\documentclass[3p,times,procedia]{elsarticle}
\usepackage{nupha_ecrc}

\volume{00}

\firstpage{1}

\journalname{Nuclear Physics A}

\runauth{Wei Li}

\jid{nupha}

\jnltitlelogo{Nuclear Physics A}

\usepackage{amssymb}
\usepackage{amsmath}

\usepackage{lineno}

\biboptions{sort&compress}

\usepackage[figuresright]{rotating}
\usepackage{xspace}

\begin{document}

\begin{frontmatter}



\dochead{XXVIth International Conference on Ultrarelativistic Nucleus-Nucleus Collisions\\ (Quark Matter 2017)}

\title{Collective flow from AA, pA to pp collisions \\\
       -- {\it {\Large Toward a unified paradigm}}}


\author{Wei Li}

\address{
Rice University, 6100 Main St, MS-315, Houston, TX 77005, USA}

\ead{davidlw@rice.edu}

\begin{abstract}

I give an overview of the latest development in understanding 
collective phenomena in high-multiplicity hadronic final state from 
relativistic nucleus-nucleus, proton-nucleus and proton-proton collisions.
Upon reviewing the experimental data and confronting them with
theoretical models, a unified paradigm in describing the observed collectivity
across all hadronic collision systems is emerging. Potential future paths toward
addressing key open questions, especially on collectivity in small systems 
(pp, pA), are discussed.

\end{abstract}


\end{frontmatter}


\vspace{-1mm}
\section{Introduction}
\label{sec:intro}

Collective phenomena have been a central theme in the research of strongly 
correlated, interacting many-body systems in nearly all fields of physics, 
from the astrophysical scale like the neutron stars to the most fundamental 
scale, such as the ``quark-gluon plasma'' (QGP). They form the ``complex frontier''
in physics, which addresses issues of the “emergence”, or emergent phenomena. 
The most important, key question to ask is what the fundamental laws are in
describing the emergent phenomena, or taking a different point of view, 
how the emergent phenomena can be understood from the fundamental forces,
if possible at all.

\begin{figure}[thb]
\center
\includegraphics[width=0.9\linewidth]{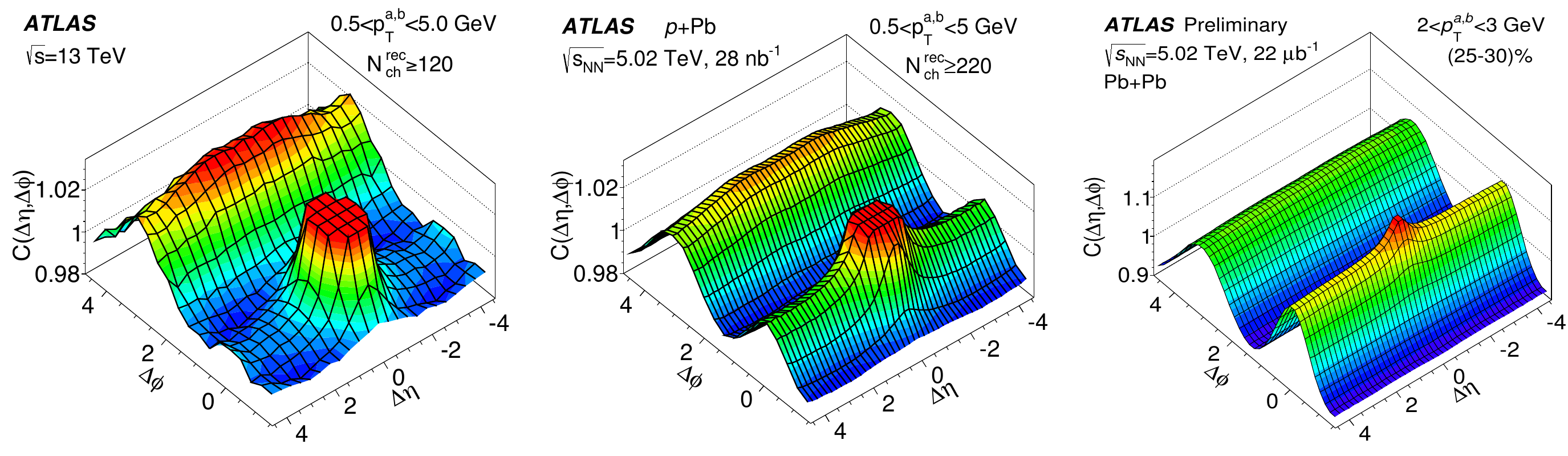}
\vspace{-3mm}
  \caption{ \label{fig:ridge} 2-D two-particle correlation functions
  in high-multiplicity pp, pPb and PbPb collisions at the LHC from ATLAS~\cite{atlas:2012fa,Aad:2015gqa}. 
   }
\end{figure}

In high-energy proton-proton (pp), proton-nucleus (pA) and nucleus-nucleus 
(AA) collisions, an emergent phenomenon of a novel anisotropy azimuthal 
correlation is observed in high-multiplicity events
~\cite{Khachatryan:2010gv,Aad:2015gqa,Khachatryan:2015lva,CMS:2012qk,alice:2012qe,atlas:2012fa,Aaij:2015qcq,Chatrchyan:2012wg,Aamodt:2011by,ATLAS:2012at}. 
As shown in Fig.~\ref{fig:ridge}, a $\cos2(\Delta\phi)$-like azimuthal structure 
is observed in two-particle $\Delta\eta$--$\Delta\phi$ correlation functions,
most notably in pPb and PbPb collisions but also in high-multiplicity pp collisions.

There are two most striking features of this novel correlation, which have
imposed strong constraints on the possible interpretations of its underlying physics mechanisms: (1)
first, this azimuthal structure is long-range in rapidity, extending over at least
5 even up to 10 units, thus known as the ``ridge''; (2) secondly, 
the correlation is found to be collective involving nearly all particles 
produced in the event. This is evidenced by the fact that the strengthen 
of elliptic flow, $v_2$, is nearly invariant when extracted by correlating no matter how many
particles at a time. Also, the $v_2$ values in pp and pPb are observed to be roughly independent of
multiplicity~\cite{Khachatryan:2015waa,Khachatryan:2016txc}, 
indicating that as more particles are incorporated into the system,
they immediately follow the same behavior of collectivity as the rest of the system.
The mass dependence of identified particle yield and $v_2$ also suggests
the production of particles from a collectively moving
source~\cite{Abelev:2013haa,ABELEV:2013wsa,Khachatryan:2016yru,Khachatryan:2014jra,Khachatryan:2016txc}. 

The collective property of the ridge indicates the presence of strong interactions 
either during the (final) stage of system evolution, or at the initial stage 
before particles are produced. The long-range nature in rapidity then
implies that the anisotropies must be rooted at an early time 
constrained by causality, $\tau_{0} \leq \tau_{f.o.} \exp\left(-\frac{1}{2}\vert y_a - y_b\vert\right)$,
where $y_{a,b}$ is the momentum-space rapidity of particles $a,b$~\cite{Dumitru:2010iy}.  
Namely, before a time-scale of $\sim$ 0.1~fm/c, an initial anisotropy 
must be present either in the position space or in the momentum space. 
For the first scenario of position-space anisotropy, strong final-state
interactions are necessary to transpose it into the final observed anisotropy in
the momentum space. This scenario includes models of hydrodynamics,
parton transport and/or escape. For the second scenario, 
momentum space anisotropies are already present
before the collision occurs via initial interactions of gluons inside
the projectile proton or nucleus. Models belonging to this category include
color glass condensate (CGC) glasma model, color-field domains and etc.
Any theoretical model proposed must fall into one scenario or the other. Otherwise, 
it is immediately ruled out by the experimental data (see a review in Ref.~\cite{Dusling:2015gta}
and references therein).

\vspace{-2mm}
\section{Collective flow in large systems -- AA}
\label{sec:large}

Based on extensive studies over the past couple of decades,  
the community has converged to the first scenario of
strong final-state interactions in describing the collective anisotropy
flow observed in large AA collision systems.
A paradigm of a nearly perfect QGP fluid formed has been established. 
Deformations present in a lumpy initial energy density profile are fully 
transposed, by nearly ideal hydrodynamics on an event-by-event basis, into 
anisotropies in the final-state particle momentum distribution. 
Other components of the model are also introduced to describe pre-equilibrium, 
freeze-out and hadronic transport stages. Fourier bases have been applied 
to characterize the final-state anisotropy flow, $f(p_{T},\eta,\phi) = N(p_{T},\eta)\displaystyle\sum_{-\infty}^{\infty}\vec{V}_{n}(p_{T},\eta)e^{-in\phi}$, with the zeroth-order term, $N(p_{T},\eta)$, providing information of the radial flow strength, 
and Fourier vector coefficients, $\vec{V}_{n}(p_{T},\eta)$ (=$v_{n}e^{in\Psi_{n}}$), 
characterizing the magnitude ($v_n$) and orientation ($\Psi_{n}$)
of flow anisotropies. The $v_n$ values extracted experimentally 
up to higher orders are found to be well described by hydrodynamic models with a very 
small value of shear viscosity to entropy density ratio ($\eta/s \lesssim 0.2$),
close to the unitarity bound on $\eta/s$ from the holographic 
principle~\cite{Kovtun:2003wp}. 


With the success of establishing the perfect QGP liquid paradigm, the field
of heavy-ion physics has entered a precision era. Many open questions related to the collective flow still
remain to be addressed, e.g.: (1) {\it How to further improve the modeling of the
initial state in 3-D and constrain better the QGP's transport properties ($\eta/s$,$\zeta/s$)
as well as their temperature dependence?} (2) {\it Do small systems like pp and pPb collisions fit into this paradigm?} 
To address these questions, the ideal goal is to obtain
fully the event-by-event information of particle spectra and all orders of
anisotropy flow harmonics by unfolding the probability density function ($p.d.f$).
Practically, experimentalists construct the ``new'' flow observables
to extract various orders of moments and cumulants of the $p.d.f$. Two main
recent directions are discussed below.

\vspace{-2mm}
\subsection{Mixed-order harmonic correlations -- $\left\lbrace\vec{V}_{n},\vec{V}_{m}\right\rbrace$}

Correlations of flow harmonic magnitudes between different orders have been studied by ALICE in PbPb collisions using a 
four-particle cumulant approach (or symmetric cumulant)~\cite{ALICE:2016kpq},
$SC(n,m)=\left<v^{2}_{n}v^{2}_{m}\right>-\left<v^{2}_{n}\right>\left<v^{2}_{m}\right>$.
An anti-correlation between $v_2$ and $v_3$ is observed, while $v_2$ and $v_4$ are positively 
correlated. The general feature of the data is captured by hydrodynamic models. As $v_2$ and $v_3$ 
are linearly proportional to the initial eccentricities, after normalizing by the $v_2$ 
and $v_3$ magnitudes, $SC(2,3)/\left(\left<v^{2}_{2}\right>\left<v^{2}_{3}\right>\right)$ is shown to be
insensitive to the medium transport properties, and thus can directly probe properties of 
initial-state geometry fluctuations, more specifically the fluctuating granularities. 

\begin{figure}[htb]
\center
\includegraphics[width=0.9\linewidth]{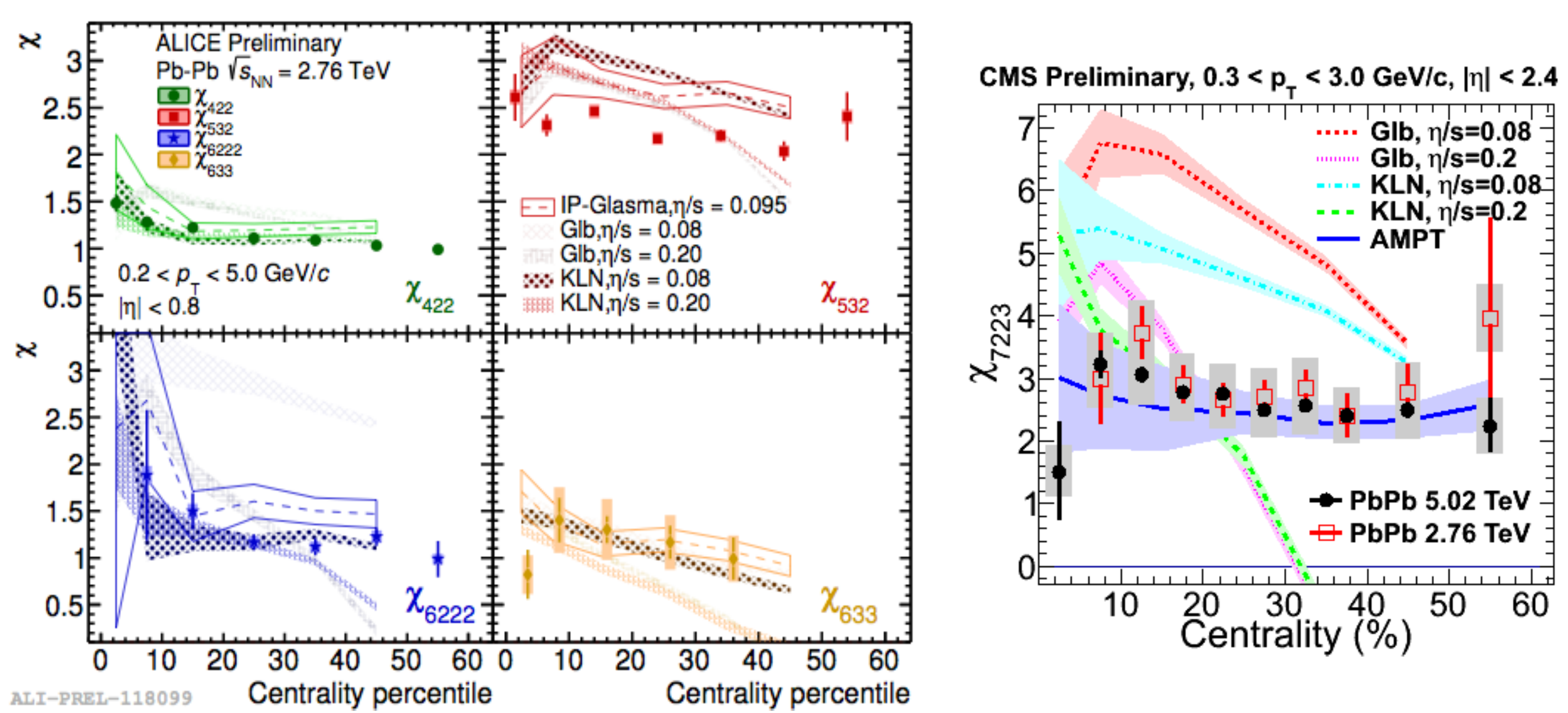}
  \vspace{-2mm}
  \caption{ \label{fig:mixedvn} 
  The non-linear response coefficients for various orders of harmonic mixings
  in PbPb collisions as a function of centrality, measured by ALICE and CMS.
  Calculations from hydrodynamic and parton transport models are also shown.
   }
\end{figure}

Correlation between $v_2$ and $v_4$ is more complicated,
showing a strong dependence on both the initial state and transport properties.
The reason has been understood as the breakdown of linear responses to the initial-state
geometry for higher-order harmonics, $v_n$ ($n \geq 4$), which receive a 
non-linear contribution from mixings of lowest-order $v_2$ and $v_3$
harmonics, as described by following equations,
\vspace{-4mm}

\begin{equation}
\vec{V}_{4}=\vec{V}_{4L}+\chi_{422}\left(\vec{V}_{2}\right)^{2},\vec{V}_{5}=\vec{V}_{5L}+\chi_{523}\vec{V}_{2}\vec{V}_{3},
\vec{V}_{6}=\vec{V}_{6L}+\chi_{6222}\left(\vec{V}_{2}\right)^{3}+\chi_{633}\left(\vec{V}_{3}\right)^{2},\vec{V}_{7}=\vec{V}_{7L}+\chi_{7223}\left(\vec{V}_{2}\right)^{2}\vec{V}_{3}.
\label{eq:mixed}
\end{equation}

\noindent The non-linear terms are sensitive to final-state dynamics of the QGP, 
as opposed to the leading terms, which are approximately linear with initial eccentricities.
As leading and non-linear terms are orthogonal to each other, they can be decomposed by
a projection of higher-order flow vector onto the lowest-order ones
and extract the so-call non-linear response coefficients, $\chi$,
for various mixing combinations~\cite{Yan:2015jma,Qian:2016fpi}.

New measurements of non-linear response coefficients are performed by the ALICE and CMS in PbPb
collisions, shown in Fig.~\ref{fig:mixedvn}, averaged over the full $p_{T}$ range 
as a function of centrality for five different mixing coefficients.
Overall, weak centrality dependence for all coefficients is observed. The data are compared 
to predictions of hydrodynamic and parton transport (AMPT) models. Impressively, the highest-order
mixing coefficient, $\chi_{7223}$, is well described by the AMPT model. The hydrodynamic model can
capture qualitative features of the data but has a strong sensitivity to both
initial-state models and $\eta/s$ values. As the harmonic mixings are mainly determined at
the freeze-out stage, these high precision new data show great promise in providing
unique constraints on the modeling of freeze-out dynamics of the QGP evolution.

\vspace{-2mm}
\subsection{Correlations of harmonics at different ($\eta$, $p_{T}$) -- $\left\lbrace\vec{V}_{n}(\eta_{1},p_{T,1}),\vec{V}_{n}(\eta_{2},p_{T,2})\right\rbrace$}

Much of studies on collective flow have previously been focusing on the midrapidity 
region, whereas the QGP evolution takes place in 3-D. Several important 
questions related to the longitudinal dynamics of QGP have not yet been well 
addressed, e.g., (1) {\it how is the initial entropy deposited in 3-D space? How 
does it fluctuate event-by-event?}; (2) {\it What is the role of the longitudinal pressure gradient?}
These issues start being explored in recent couple of years by studying 
correlation of flow harmonic vectors at different rapidities and 
transverse momentum.

\begin{figure}[thb]
\center
\includegraphics[width=0.9\linewidth]{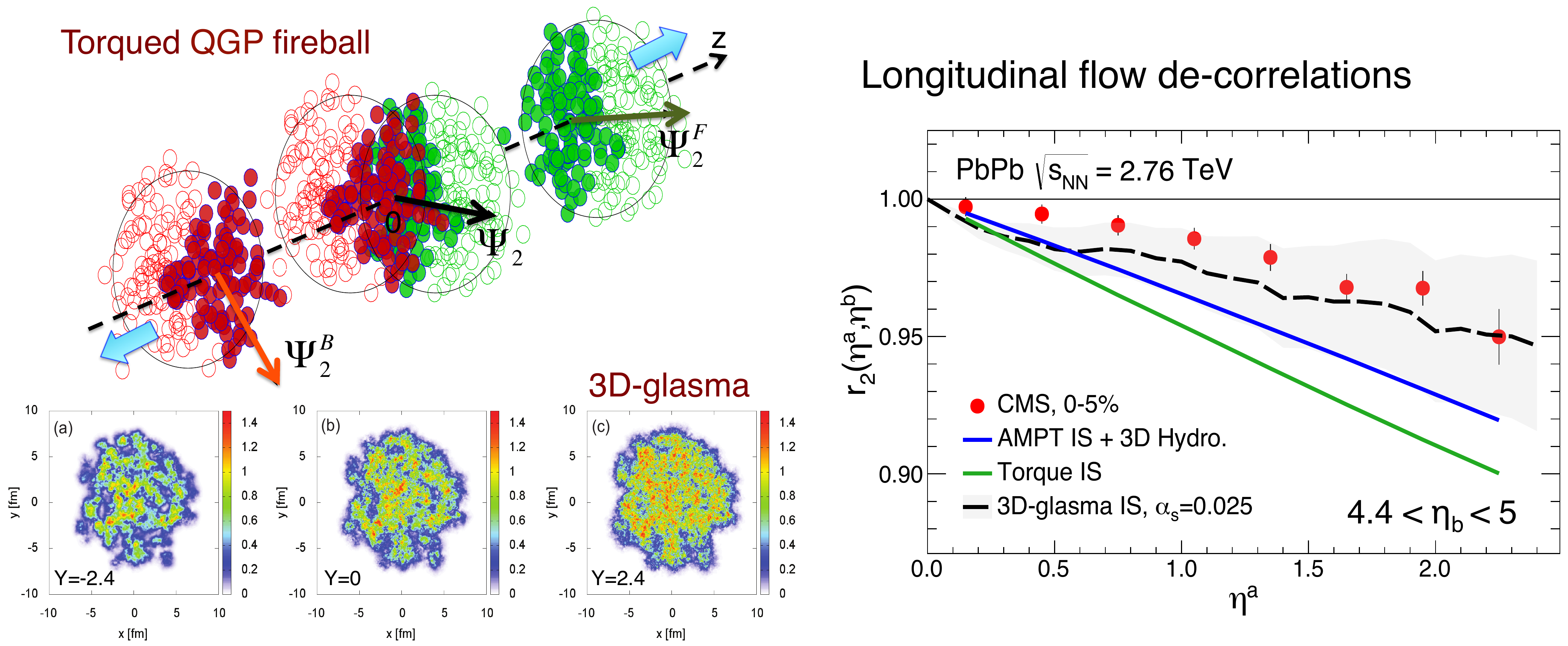}
\vspace{-2mm}
  \caption{ \label{fig:vnlong} Left: illustration of flow event plane
  de-correlations as a function of rapidity in the wounded nucleon picture 
  (or ``torqued QGP fireball'')~\cite{Bozek:2010vz} 
  and 3-D color glass condensate model~\cite{Schenke:2016ksl}.
  Right: measurement of elliptic flow de-correlations 
  as a function of pseudorapidity in 0-5\% central PbPb collisions at 2.76~TeV 
  from CMS~\cite{Khachatryan:2015oea}, with comparison to 
  theoretical calculations~\cite{Bozek:2010vz,Schenke:2016ksl,Pang:2015zrq}.
   }
\end{figure}

Novel rapidity-dependent event plane twist or de-correlation has been
predicted, as illustrated in Fig.~\ref{fig:vnlong} (left). 
In the picture of wounded nucleon model~\cite{Bozek:2010vz}, particles produced 
at midrapidity receive about equal contribution from participants of both nuclei. 
However, if going to forward rapidity region, particles will be predominantly produced
from one of the projectile nuclei. As a result, the flow orientation angle (or event plane) 
at forward and backward rapidities can be slightly twisted event-by-event, 
creating a torqued QGP along rapidity direction. Additionally, in the CGC glasma 
model~\cite{Schenke:2016ksl}, fluctuating granularity of the gluon field 
is rapidity dependent. As moving toward
larger rapidity and smaller x value, the initial configuration of gluon fields 
tends to become smoother. Both effects of participant nucleon and gluon field fluctuations 
can induce rapidity-correlated flow fluctuations.

The rapidity-dependent flow de-correlations have been observed by
CMS by constructing a ratio of flow correlators between two
flow vectors measured at different rapidity regions, 
$r_{n} \equiv \frac{\left<\vec{V}_{n}(-\eta^{a})\vec{V}^{*}_{n}(\eta^{b})\right>}{\left<\vec{V}_{n}(\eta^{a})\vec{V}^{*}_{n}(\eta^{b})\right>}$, designed to approximate the 
de-correlation between two event plane angles separated by a gap of 
$2\eta_{a}$, $\left<\cos n\left[\Psi_{n}(\eta^{a})-\Psi_{n}(-\eta^{a})\right]\right>$,
as shown in Fig.~\ref{fig:vnlong} (right) for elliptic flow
in 0--5\% central PbPb collisions. The data are compared to 
several models of initial states including torqued QGP model, AMPT initial state 
followed by a 3-D hydrodynamics and 3-D CGC glasma model, which all 
qualitatively reproduce the data. It is worth noting that almost all
the rapidity de-correlation effect is determined from the initial state,
addition of 3-D hydrodynamic evolution is found to have little impact 
on the $r_{n}$ ratio. This underlines the importance to incorporate
a rapidity dependent modeling of initial-state fluctuations in hydrodynamic
calculations.

New studies on flow de-correlations in rapidity in PbPb collisions are performed 
by ATLAS, using a new ratio of correlators among four flow vectors,
$R_{n} \equiv \frac{\left<\vec{V}_{n}(-\eta^{a})\vec{V}_{n}(-\eta^{b})\vec{V}^{*}_{n}(\eta^{a})\vec{V}^{*}_{n}(\eta^{b})\right>}{\left<\vec{V}_{n}(-\eta^{a})\vec{V}_{n}(\eta^{b})\vec{V}^{*}_{n}(-\eta^{a})\vec{V}^{*}_{n}(\eta^{b})\right>}$. The $R_{n}$ ratio has the advantage of 
being more sensitive to the effect of event plane angle twist, while
the originally proposed $r_{n}$ ratio is sensitive to both
event plane and participant eccentricity fluctuations in rapidity.
Comparing the $r_{n}$ and $R_{n}$ data from ATLAS in PbPb collisions,
one can conclude that event plane twist accounts for about
50\% of de-correlation effect previously observed in $r_n$.

Rapidity dependence of flow harmonics has been proposed as a means 
to probe the temperature dependence of $\eta/s$ value as temperature of the QGP
is expected to be rapidity dependent~\cite{Denicol:2015nhu}. Therefore, it is crucial to 
have a clear understanding of rapidity-correlated initial-state 
effect, which has significant contributions to the observed rapidity dependence of flow harmonics.

\vspace{-2mm}
\section{Collective flow in small systems -- pp and pA}
\label{sec:small}

\begin{figure}[bht]
\center
\includegraphics[width=0.8\linewidth]{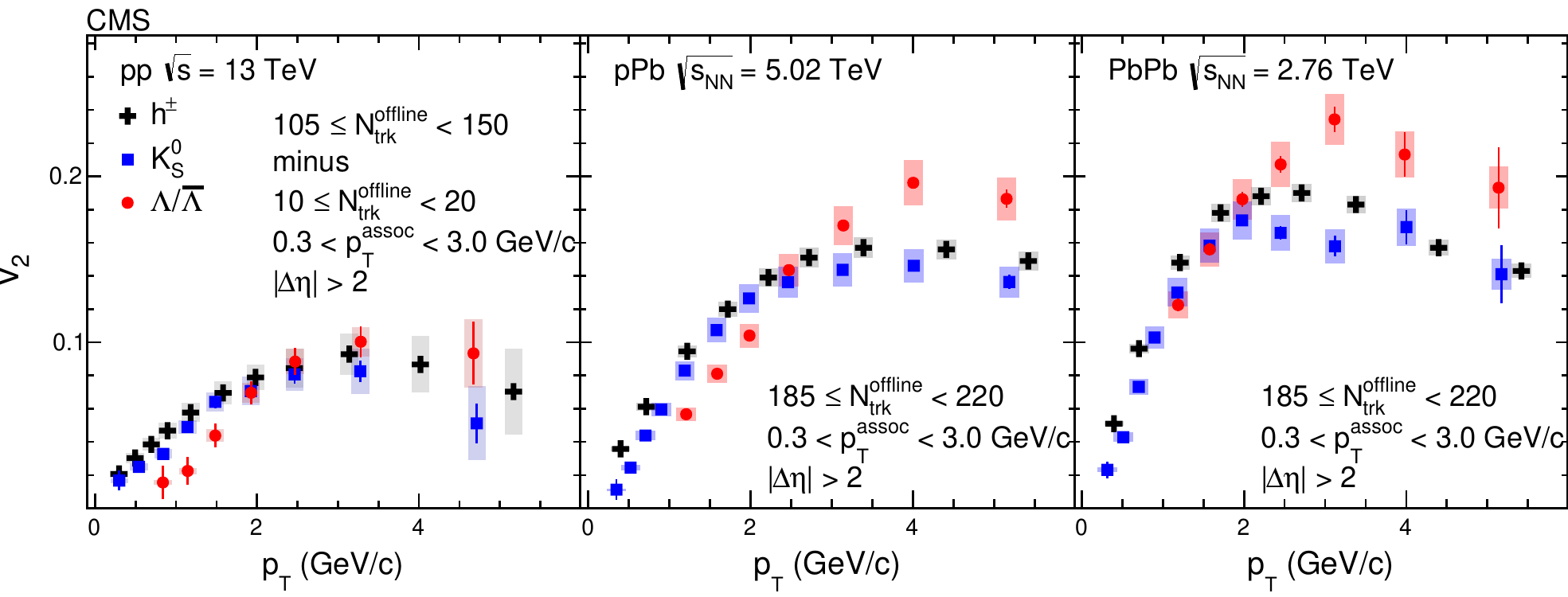}
\includegraphics[width=0.8\linewidth]{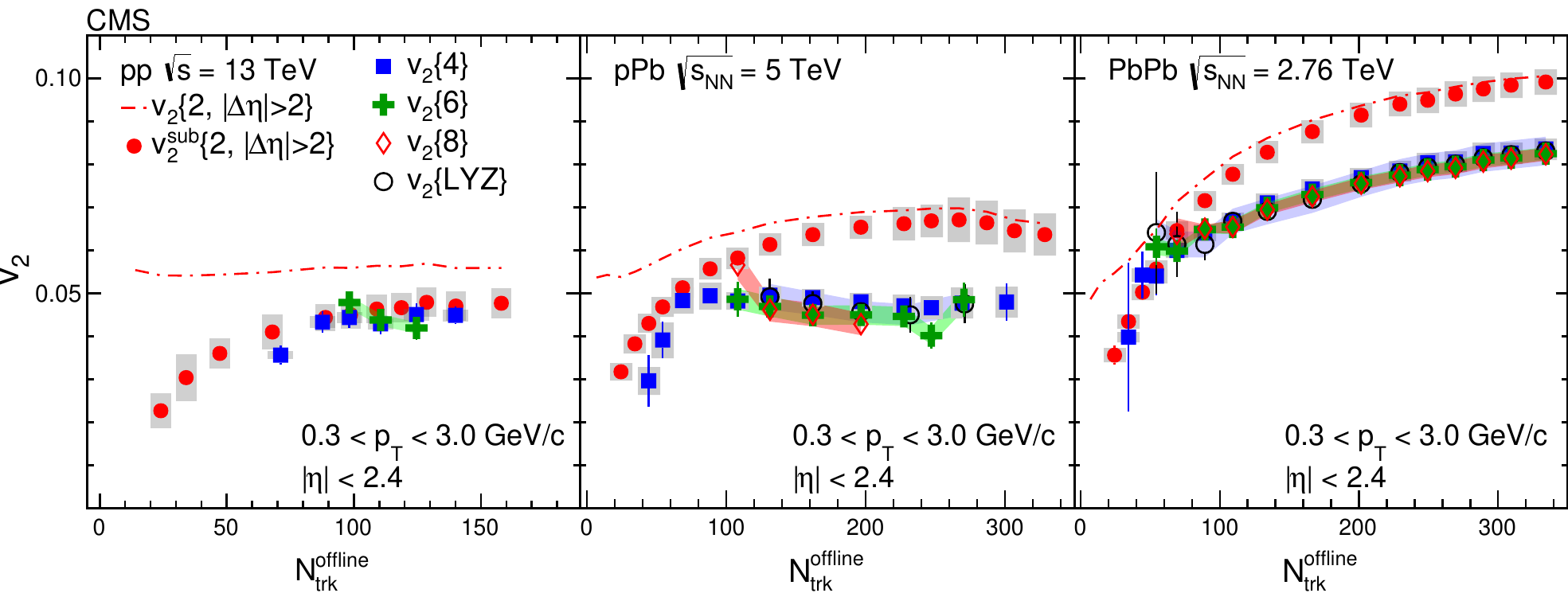}
\vspace{-2mm}
  \caption{ \label{fig:vnall} The $v_2$ data measured in pp, pPb and PbPb 
  collisions at the LHC energies by CMS as a function of $p_{T}$ for charged particles,
  $K_{0}^{s}$ and $\Lambda$ particles at high multiplicities from 
  two-particle correlations~\cite{Khachatryan:2014jra,Khachatryan:2016txc} 
  (top), and as a function of multiplicity for charged particles averaged over 
  $0.3<p_{T}<3$~GeV/c from two- and multi-particle correlations~\cite{Khachatryan:2016txc} (bottom).
   }
\end{figure}

A big question under intense debate in the field of heavy-ion physics
is: how small a QGP fluid system can be in size? In general, hydrodynamics is
applicable when the characteristic system size is much larger than its interaction
mean free path, $L \gg \lambda_{m.f.p.}$, where the mean free path is inversely 
related to the system temperature and coupling strengthen, $\lambda_{m.f.p.} \sim \frac{1}{g^{4}T}$. 
In the case of strong coupling of the order of 1, hydrodynamics requires a condition of
$LT \gg 1$. On the other hand, in the limit of extremely strong coupling as for the
holographic principle, this criteria could be significantly loosened to $LT \sim 1$
so that a QGP fluid may be realized with a much smaller size ($\sim 1/T$) at a given temperature.
Therefore, the question on how small a QGP fluid can be has important implications 
to the most fundamental property of the QGP medium. 

To emphasize again, it is $LT$ that determines the system's
fluid behavior, instead of just the absolute size. So
what is the corresponding experimental condition then? As entropy density, $s$, 
scales as $T^{3}$ for a thermalized QGP system, and also $s$ is approximately 
proportional to event multiplicity, $N_{\rm trk}$ over $L^{3}$, a qualitative
relation can be derived that $LT \sim (N_{\rm trk})^{\frac{1}{3}}$. Therefore, 
the most relevant question to ask may not be about absolute size of the 
system but, instead, what is the smallest multiplicity or total entropy the system
has to produce to exhibit hydrodynamic behavior. Lots of experimental evidence 
also suggest that total event multiplicity does seem to play a special role
in driving the collective effects of produced particles.

\begin{figure}[thb]
\center
\includegraphics[width=0.425\linewidth]{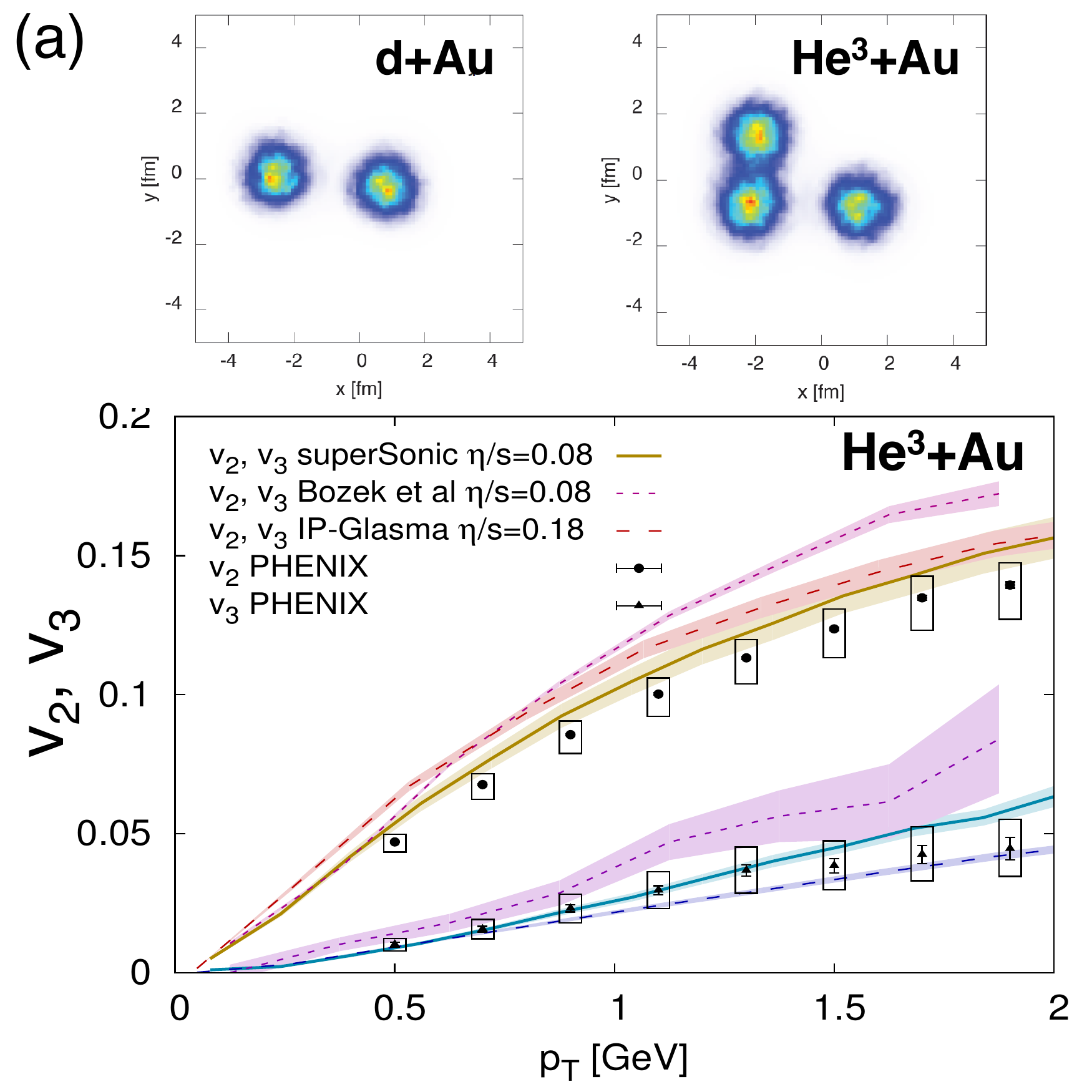}
\includegraphics[width=0.425\linewidth]{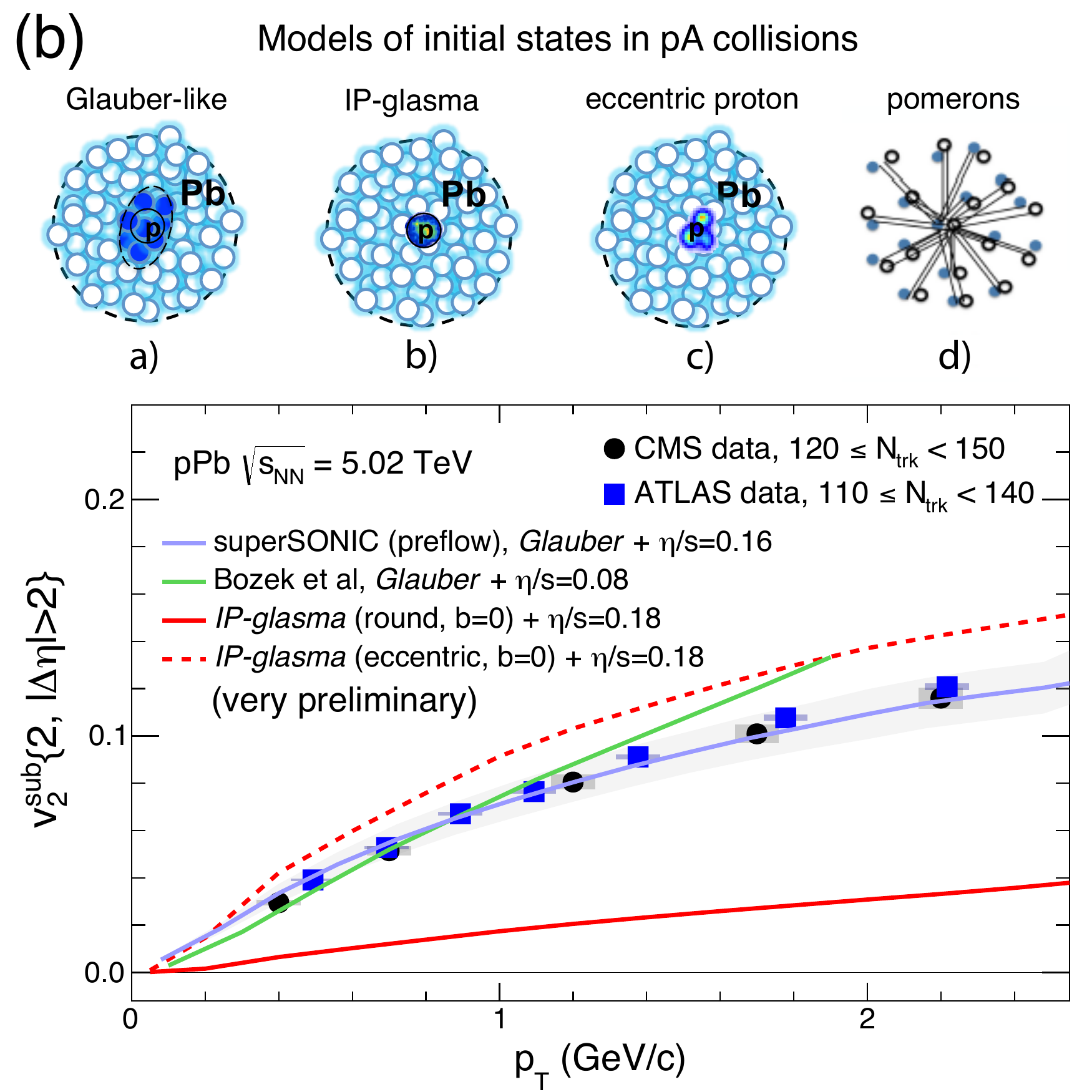}
\vspace{-2mm}
  \caption{ \label{fig:vnsmall} Left: The $v_2$ and $v_3$ data measured in 0--5\% central 
  He$^{3}$Au collisions at 200 GeV by PHENIX as a function of $p_{T}$~\cite{Adare:2015ctn}. Right: 
  the $v_2$ data measured in high-multiplicity pPb collisions at 5.02 TeV 
  by CMS and ATLAS as a function of $p_{T}$~\cite{Chatrchyan:2013nka,atlas:2012fa}. 
  Comparison to hydrodynamic
  model calculations is also shown with different modeling of initial-state
  fluctuations as illustrated in the cartoons.
   }
\end{figure}

Several key features of collectivity have recently been observed in high-multiplicity pp
collisions as well, similar to those in pPb and PbPb collisions. 
These include, as shown in Fig.~\ref{fig:vnall} for data from CMS, mass ordering
of $v_2$ (comparing charged particles, $K_{s}^{0}$ and $\Lambda$), multi-particle cumulant $v_2$ 
($v_{2}\{4\} \approx v_{2}\{6\}$) that is independent of multiplicity, and mass dependent of
identified particle spectra (not shown). Experimental observation of collective behavior across all hadronic 
collision systems with high-multiplicity final state has been established, although this
does not necessarily imply a hydrodynamic origin of collectivity.

Progress has also been made recently for the scenario of initial interaction models. By incorporating
the Lund string model in PYTHIA to fragment gluons into final-state hadrons, 
the CGC glasma model is able to make direct quantitative comparisons to experimental 
observables for the first time. For instance of Ref.~\cite{Schenke:2016lrs}, mass ordering of $v_2$
in high-multiplicity pp is reproduced by the CGC glasma model. Next important step is to examine
other observables relevant to collectivity such as multi-particle cumulants. 

The key to further differentiate between the initial- and final-state interaction scenarios 
(both may give rise to collective particle correlations) is to investigate the connection
between final-state collective anisotropies and initial-state geometry (or eccentricity),
which has been well established in AA but not yet for small systems. 
Smallness in absolute size is not the limitation 
in hydrodynamics. This has been convincingly demonstrated in a series of 
geometry-controlled experiments on small systems at RHIC by colliding light ions 
like dAu~\cite{Adare:2014keg} and He$^{3}$Au~\cite{Adare:2015ctn}. 
For those systems, where each nucleus contains at least two nucleons, 
initial eccentricities are still largely determined by the position of wounded nucleons, 
which has been well understood in the Glauber picture. As illustrated in Fig.~\ref{fig:vnsmall}a
(top), a dAu or He$^{3}$Au collision has the configuration of two-blob
or three-blob deposited energy in its initial state. Later on, each blob will expand and
generate a shock wave. Collisions of shock waves will lead to $v_2$ and $v_3$ anisotropy.
Fig.~\ref{fig:vnsmall}a (bottom) shows the $v_2$ and $v_3$ data
in 0--5\% central He$^{3}$Au collisions from PHENIX~\cite{Adare:2015ctn}. The data are in good agreement with
hydrodynamic calculations using either Glauber or IP-glasma initial state, providing 
clear evidence for the applicability of hydrodynamics in system as small as a couple of $fm$ in size.

However, when moving to pA (and pp) system, the agreement among different models and data
does not hold anymore, as shown in Fig.~\ref{fig:vnsmall}b (bottom) for $v_2$ data in pPb
collisions from CMS and ATLAS. The main issue here is that if one of the projectiles is a single nucleon,
the initial geometry of the overlap region is highly sensitive to the details of 
event-by-event shape of a nucleon, which is poorly known. Different approaches in modeling
the initial state in pA collisions are demonstrated in Fig.~\ref{fig:vnsmall}b (top).
The IP-glasma model significantly
underestimates the $v_2$ data in pPb because the geometry of the overlap zone is largely 
determined by the proton's shape, which is spherical. In the Glauber
model, it assumes that the entire nucleon of those wounded ones would contribute to the 
initial geometry. This leads to a much larger eccentricity because of more degree of freedom for 
fluctuations. Modeling of event-by-event proton shape fluctuations in the framework of CGC model
has been recently attempted and shown promising results in generating a much larger final-state $v_2$ value,
closer to the experimental data, as one can see in Fig.~\ref{fig:vnsmall}b (bottom) for
IP-glasma model with an eccentric proton shape. Therefore, small systems like pA and pp collisions
provide us an exciting opportunity of imaging subnucleonic-scale quantum fluctuations over yoctoseconds 
for the first time, a unique opportunity that does not exist in large systems.

\begin{figure}[thb]
\center
\includegraphics[width=0.8\linewidth]{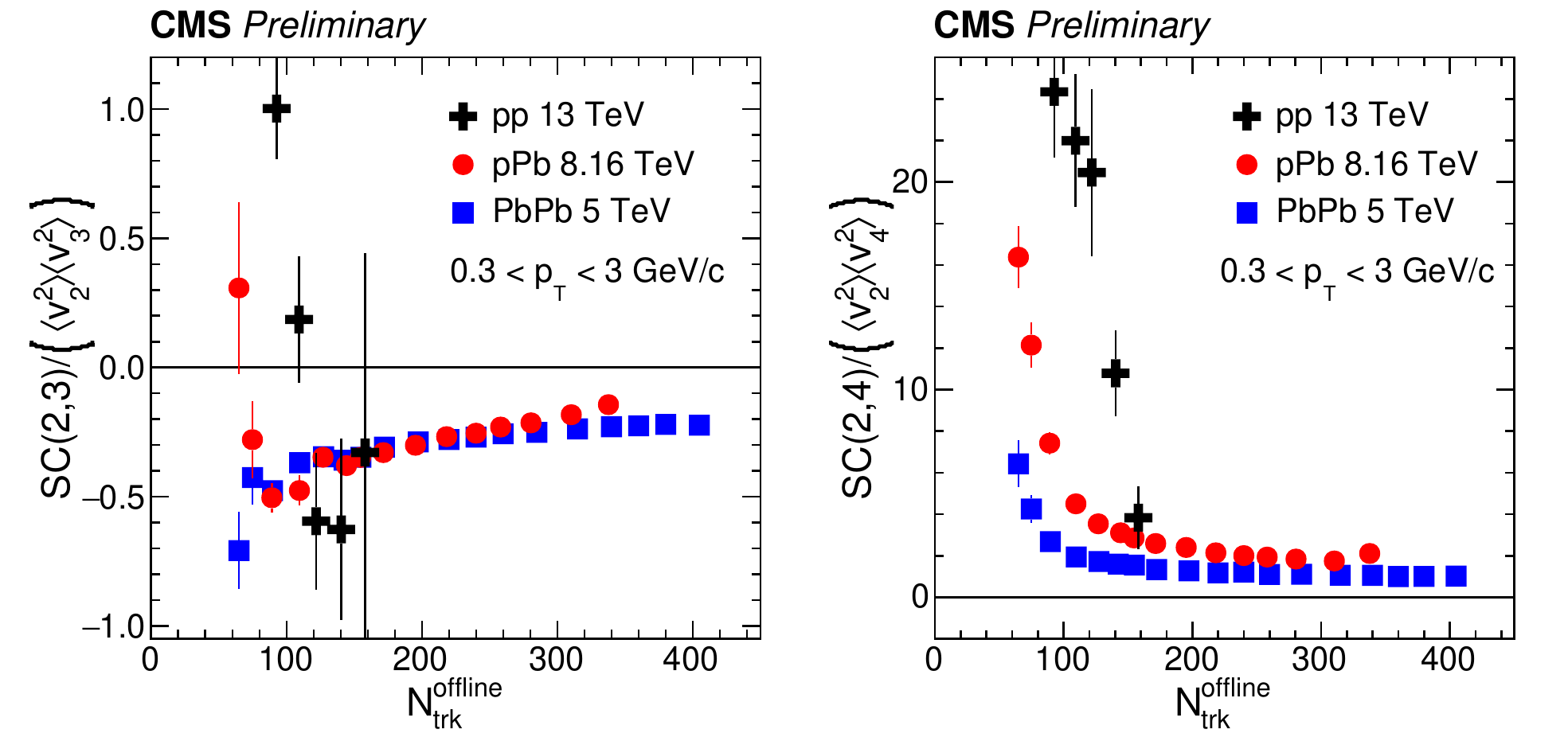}
\vspace{-2mm}
  \caption{ \label{fig:sccms} Correlations of $v_{2}$--$v_{3}$ and $v_{2}$--$v_{4}$
  measured using the symmetric cumulants, normalized by $\left<v^{2}_{2}\right>\left<v^{2}_{3}\right>$ 
  and $\left<v^{2}_{2}\right>\left<v^{2}_{4}\right>$, as a function of multiplicity
  in pp, pPb and PbPb collisions at the LHC energies from CMS.
   }
\end{figure}

To probe the connection to initial-state fluctuations in pp and pA collisions,
new flow observables can be employed, as learned from the studies of AA collisions.
As an example, new measurements of $v_2$--$v_3$ and $v_2$--$v_4$ correlations
in pp and pPb collisions, using the symmetric cumulant method, have been performed
by CMS at the LHC. A clear anti-correlation between $v_2$ and $v_3$ is observed,
similar to that in PbPb collisions. This feature can be naturally accounted for in 
hydrodynamic picture by the initial-state geometry. It remains to be seen
whether such correlations of anisotropy harmonics are also present in initial
interaction models, and what sign of the correlation is. In Fig.~\ref{fig:sccms},
normalized symmetric cumulants are compared among pp, pPb and PbPb systems
as a function of multiplicity measured by CMS. Particularly for $SC(2,3)$, which
is mainly sensitive to the initial state, the magnitude of the correlation is found
to be quantitatively similar between pPb and PbPb, and maybe even pp. 
This may indicate again a similar origin of the observed correlations
across all systems, possible related to the fluctuating granularity of the QGP's 
initial-state geometry.

\vspace{-2mm}
\section{Outlook -- Qpen questions in small systems}

As an outlook, a few important open questions related to
collectivity in small systems are discussed below as a guide to
possible future directions to be pursued by the community.

{\it Does the collectivity turn off at low multiplicity?} The trend of $v_2$ 
in pp collisions from two-particle correlations toward the low multiplicity region
still has large uncertainties (constant from ATLAS and decreasing from CMS). 
As argued earlier, multiplicity is the quantity that drives the collective flow 
in the hydrodynamic picture. So a $v_2$ of hydrodynamic origin should decrease 
as multiplicity becomes sufficiently low. Intuitively, hydrodynamic collectivity 
develops over time. Low multiplicity events tend to have a shorter lifetime,
thus developing a smaller $v_2$ anisotropy. If a constant trend of $v_2$ at low multiplicities is confirmed, 
an alternative interpretation may be needed. More detailed information 
is necessary for resolving this dispute. For example, identified particle spectra and 
$v_2$ data in pp collisions suggest that mass ordering effect tends to
vanish as multiplicity decreases, consistent with the hydrodynamic scenario. In terms of $v_2$
from multi-particle cumulants, a possible path forward is to implement rapidity gaps between particles 
as for two-particle correlations so that short-range non-collective correlations would be suppressed.

{\it Can we observe jet quenching in small systems?} If the observed collectivity in small systems is 
suggestive of strong final-state interactions, quenching of high $p_{T}$ jets should also be present. 
The energy loss from a pQCD approach is expected to follow a dependence on $\hat{q}$
and path length (or system size), $L$, as $\Delta E \sim \alpha_s(T)\hat{q}(T)L^{2}$. 
As $\hat{q}$ is expected to go as $T^{3}$, one arrives at $\Delta E \sim T^{3}L^{2}$.
If comparing at similar multiplicities, a smaller (in $L$) system possesses a higher entropy density 
or temperature ($T$). Therefore, sizeable parton energy loss should also be expected for 
high-multiplicity pp and pA systems, comparable to that for peripheral AA collisions. In search for
jet quenching in small systems, the main complication lies in the non-trivial correlation between 
underlying event multiplicity and hard probes, making it difficult to construct a proper in-vacuum reference.
With a high-luminosity pPb run delivered by the LHC in 2016, a promising measurement to pursue would be
the $v_2$ at very high $p_{T}$ ($\gtrsim 10$~GeV/c) using multi-particle cumulants,
which is related to the path length dependence of energy loss.

{\it Does the collectivity extend to non-hadronic collisions (e.g., ep, eA, UPC and e$^{+}$e$^{-}$)? What are the fundamental
requirements for creating a microscopic fluid?} While high-multiplicity final state seems to be a necessary condition,
initial colliding projectiles may not have to be hadronic but universal also for electromagnetic probes like electrons and photons.
If collective flow behavior can be observed in ep, eA, UPC and even e$^{+}$e$^{-}$ collisions with
high-multiplicity hadronic final state, it will open a new exciting avenue of research in many-body QCD system.
After all, once the QCD vacuum is excited by collision of strong fields to generate sufficient initial entropy,
it may then flow collectively like a perfect fluid.

\vspace{-2mm}
\section{Summary}
\label{sec:summary}
In summary, clear evidence of long-range collective phenomena has been observed, 
and it is universal in all high-multiplicity hadronic collisions. Possible interpretations 
have to fall into only two scenarios distinguished by whether the interactions are at 
final or initial stage. There is a wide consensus that collectivity in AA collisions
is driven by strong final-state interactions of a fluid-like QGP medium. However,
collectivity seen in small systems raised debates on whether the ``perfect'' fluid paradigm
is still valid or not. The key to address this question is to seek for the 
connection to the initial-state geometry. New direction in studying new flow observables, such as flow
harmonic correlations, in small systems shows good promises and may provide 
unique opportunities of probing subnucleonic-scale quantum
fluctuations for the first time. Future paths toward addressing several
open questions on collectivity in small systems are discussed.

\vspace{-2mm}
\section*{Acknowledgement}
WL acknowledges funding from a DOE Office of Science Early Career Award (Contract No. DE-SC0012185), from
the Welch Foundation (Grant No. C-1845) and from an Alfred P. Sloan Research Fellowship (No. FR-2015-65911).





\vspace{-2mm}
\bibliographystyle{elsarticle-num}
\bibliography{nupha-weili}

\begin{thebibliography}{10}
\expandafter\ifx\csname url\endcsname\relax
  \def\url#1{\texttt{#1}}\fi
\expandafter\ifx\csname urlprefix\endcsname\relax\def\urlprefix{URL }\fi
\expandafter\ifx\csname href\endcsname\relax
  \def\href#1#2{#2} \def\path#1{#1}\fi

\bibitem{atlas:2012fa}
G.~Aad, et~al., Phys. Rev. Lett. 110 (2013) 182302.
\newblock \href {http://arxiv.org/abs/1212.5198} {\path{arXiv:1212.5198}}.

\bibitem{Aad:2015gqa}
G.~Aad, et~al., Phys. Rev. Lett. 116 (2016) 172301.
\newblock \href {http://arxiv.org/abs/1509.04776} {\path{arXiv:1509.04776}}.

\bibitem{Khachatryan:2010gv}
V.~Khachatryan, et~al., JHEP 09 (2010) 091.
\newblock \href {http://arxiv.org/abs/1009.4122} {\path{arXiv:1009.4122}}.

\bibitem{Khachatryan:2015lva}
V.~Khachatryan, et~al., Phys. Rev. Lett. 116 (2016) 172302.
\newblock \href {http://arxiv.org/abs/1510.03068} {\path{arXiv:1510.03068}}.

\bibitem{CMS:2012qk}
S.~Chatrchyan, et~al., Phys. Lett. B 718 (2013) 795.
\newblock \href {http://arxiv.org/abs/1210.5482} {\path{arXiv:1210.5482}}.

\bibitem{alice:2012qe}
B.~Abelev, et~al., Phys. Lett. B 719 (2013) 29.
\newblock \href {http://arxiv.org/abs/1212.2001} {\path{arXiv:1212.2001}}.

\bibitem{Aaij:2015qcq}
R.~Aaij, et~al., Phys. Lett. B 762 (2016) 473.
\newblock \href {http://arxiv.org/abs/1512.00439} {\path{arXiv:1512.00439}}.

\bibitem{Chatrchyan:2012wg}
S.~Chatrchyan, et~al., Eur. Phys. J. C 72 (2012) 2012.
\newblock \href {http://arxiv.org/abs/1201.3158} {\path{arXiv:1201.3158}}.

\bibitem{Aamodt:2011by}
K.~Aamodt, et~al., Phys. Lett. B 708 (2012) 249.
\newblock \href {http://arxiv.org/abs/1109.2501} {\path{arXiv:1109.2501}}.

\bibitem{ATLAS:2012at}
G.~Aad, et~al., Phys. Rev. C 86 (2012) 014907.
\newblock \href {http://arxiv.org/abs/1203.3087} {\path{arXiv:1203.3087}}.

\bibitem{Khachatryan:2015waa}
V.~Khachatryan, et~al., Phys. Rev. Lett. 115 (2015) 012301.

\bibitem{Khachatryan:2016txc}
V.~Khachatryan, et~al., Phys. Lett. B 765 (2017) 193.
\newblock \href {http://arxiv.org/abs/1606.06198} {\path{arXiv:1606.06198}}.

\bibitem{Abelev:2013haa}
B.~B. Abelev, et~al., Phys. Lett. B 728 (2014) 25.
\newblock \href {http://arxiv.org/abs/1307.6796} {\path{arXiv:1307.6796}}.

\bibitem{ABELEV:2013wsa}
B.~B. Abelev, et~al., Phys. Lett. B 726 (2013) 164.
\newblock \href {http://arxiv.org/abs/1307.3237} {\path{arXiv:1307.3237}}.

\bibitem{Khachatryan:2016yru}
V.~Khachatryan, et~al., Phys. Lett. B 768 (2017) 103.
\newblock \href {http://arxiv.org/abs/1605.06699} {\path{arXiv:1605.06699}}.

\bibitem{Khachatryan:2014jra}
V.~Khachatryan, et~al., Phys. Lett. B 742 (2015) 200.
\newblock \href {http://arxiv.org/abs/1409.3392} {\path{arXiv:1409.3392}}.

\bibitem{Dumitru:2010iy}
A.~Dumitru, K.~Dusling, F.~Gelis, J.~Jalilian-Marian, T.~Lappi, R.~Venugopalan,
  Phys. Lett. B 697 (2011) 21.
\newblock \href {http://arxiv.org/abs/1009.5295} {\path{arXiv:1009.5295}}.

\bibitem{Dusling:2015gta}
K.~Dusling, W.~Li, B.~Schenke, Int. J. Mod. Phys. E 25 (2016) 1630002.
\newblock \href {http://arxiv.org/abs/1509.07939} {\path{arXiv:1509.07939}}.

\bibitem{Kovtun:2003wp}
P.~Kovtun, D.~T. Son, A.~O. Starinets, JHEP 10 (2003) 064.
\newblock \href {http://arxiv.org/abs/hep-th/0309213}
  {\path{arXiv:hep-th/0309213}}.

\bibitem{ALICE:2016kpq}
J.~Adam, et~al., Phys. Rev. Lett. 117 (2016) 182301.
\newblock \href {http://arxiv.org/abs/1604.07663} {\path{arXiv:1604.07663}}.

\bibitem{Yan:2015jma}
L.~Yan, J.-Y. Ollitrault, Phys. Lett. B 744 (2015) 82.
\newblock \href {http://arxiv.org/abs/1502.02502} {\path{arXiv:1502.02502}}.

\bibitem{Qian:2016fpi}
J.~Qian, U.~W. Heinz, J.~Liu, Phys. Rev. C 93 (2016) 064901.
\newblock \href {http://arxiv.org/abs/1602.02813} {\path{arXiv:1602.02813}}.

\bibitem{Bozek:2010vz}
P.~Bozek, W.~Broniowski, J.~Moreira, Phys. Rev. C 83 (2011) 034911.
\newblock \href {http://arxiv.org/abs/1011.3354} {\path{arXiv:1011.3354}}.

\bibitem{Schenke:2016ksl}
B.~Schenke, S.~Schlichting, Phys. Rev. C 94 (2016) 044907.
\newblock \href {http://arxiv.org/abs/1605.07158} {\path{arXiv:1605.07158}}.

\bibitem{Khachatryan:2015oea}
V.~Khachatryan, et~al., Phys. Rev. C 92 (2015) 034911.
\newblock \href {http://arxiv.org/abs/1503.01692} {\path{arXiv:1503.01692}}.

\bibitem{Pang:2015zrq}
L.-G. Pang, et~al., Eur. Phys. J. A 52 (2016) 97.
\newblock \href {http://arxiv.org/abs/1511.04131} {\path{arXiv:1511.04131}}.

\bibitem{Denicol:2015nhu}
G.~Denicol, A.~Monnai, B.~Schenke, Phys. Rev. Lett. 116 (2016) 212301.
\newblock \href {http://arxiv.org/abs/1512.01538} {\path{arXiv:1512.01538}}.

\bibitem{Adare:2015ctn}
A.~Adare, et~al., Phys. Rev. Lett. 115 (2015) 142301.
\newblock \href {http://arxiv.org/abs/1507.06273} {\path{arXiv:1507.06273}}.

\bibitem{Chatrchyan:2013nka}
S.~Chatrchyan, et~al., Phys. Lett. B 724 (2013) 213.
\newblock \href {http://arxiv.org/abs/1305.0609} {\path{arXiv:1305.0609}}.

\bibitem{Schenke:2016lrs}
B.~Schenke, et~al., Phys. Rev. Lett. 117 (2016) 162301.
\newblock \href {http://arxiv.org/abs/1607.02496} {\path{arXiv:1607.02496}}.

\bibitem{Adare:2014keg}
A.~Adare, et~al., Phys. Rev. Lett. 114 (2015) 192301.
\newblock \href {http://arxiv.org/abs/1404.7461} {\path{arXiv:1404.7461}}.

\end{thebibliography}







\end{document}